\documentclass[fleqn,usenatbib]{mnras}
\usepackage{newtxtext,newtxmath}
\usepackage[T1]{fontenc}
\usepackage{ae,aecompl}
\usepackage{graphicx}	
\usepackage{amsmath}	
\usepackage{amssymb}	%
\usepackage[a4paper]{geometry}
\usepackage{lmodern}

\title[Signatures of quiet Sun reconnection events.]{Signatures of quiet Sun reconnection events in Ca~{\sc ii}, H$\alpha$ and Fe~{\sc i}.}

\author[J. Shetye et al.]{
J. Shetye$^{1,4}$ \thanks{E-mail: j.shetye@warwick.ac.uk (JS)}
S. Shelyag$^{2}$
A.L. Reid$^{3}$
E. Scullion$^{2}$ 
J.G. Doyle$^{4}$
and
T.D. Arber$^{1}$
\\
$^1$Centre for Fusion, Space and Astrophysics, Department of Physics, University of Warwick, CV4 7AL \\
$^2$Department of Mathematics \& Information Sciences, Northumbria University, Newcastle Upon Tyne, NE1 8ST, UK\\
$^3$Astrophysics Research Centre, School of Mathematics and Physics, Queen's University Belfast, BT7~1NN, N. Ireland\\
$^4$Armagh Observatory and Planetarium, College Hill, Armagh BT61 9DG, N. Ireland\\
}

\date{Accepted XXX. Received YYY; in original form ZZZ}

\pubyear{2017}

\begin{document}

\maketitle

\begin{abstract}
We use observations of quiet Sun (QS) regions in the H$\alpha$ 6563 \AA, Ca~{\sc ii} 8542 \AA\, and Fe~{\sc i} 6302 \AA\ lines. We observe brightenings in the wings of the H$\alpha$ and Ca~{\sc ii} combined with observations of the interacting magnetic concentrations observed in the Stokes signals of Fe~{\sc i}. These brightenings are similar to Ellerman bombs (EBs), i.e. impulsive bursts in the wings of the Balmer lines which leave the line cores unaffected. Such enhancements suggest that these events have similar formation mechanisms to the classical EBs found in active regions, with the reduced intensity enhancements found in the QS regions due to a weaker feeding magnetic flux. The observations also show that the quiet Sun Ellerman bombs (QSEBs) are formed at a higher height in the upper photosphere than the photospheric continuum level. Using simulations, we investigate the formation mechanism associated with the events and suggest that these events are driven by the interaction of magnetic field-lines in the upper photospheric regions. The results of the simulation are in agreement with observations when comparing the light-curves, and in most cases we found that the peak in the Ca~{\sc ii} 8542 \AA\ wing occurred before the peak in H$\alpha$ wing. Moreover, in some cases, the line profiles observed in Ca~{\sc ii} are asymmetrical with a raised core profile. The source of heating in these events is shown by the MURaM simulations and is suggested to occur 430 km above the photosphere.
\end{abstract}


\section{Introduction}
Ellerman Bombs (EBs) are prominent small--scale brightenings best observed in the far wings of H$\alpha$. They were first reported by \citet{1917ApJ....46..298E} as hydrogen bombs and were termed Ellerman bombs by \cite{1960PNAS...46..165M}, while \citet{1956Obs....76..241S} termed them moustaches. They appear with a flame-like morphology, are 1000--2000 km in length and have vertical velocities of around 1 km s$^{-1}$ with durations of 10 -- 15 minutes \citep{1987SoPh..108..227Z,2002ApJ...575..506G}.  EBs are generally observed near regions with relatively high concentrations of magnetic field, such as emerging flux regions and the penumbrae of sunspots \citep[and references therein]{,2007SoPh..246...89I,2008ApJ...684..736W,2011ApJ...736...71W,2013JPhCS.440a2007R,2013ApJ...774...32V,2015ApJ...798...19N,2015ApJ...805...64R,2016ApJ...823..110R}. Magnetic field configuration occurring in the photosphere dictates the morphology of the EBs \citep{2002ApJ...575..506G,2013ApJ...774...32V,2013ApJ...779..125N,2015ApJ...805...64R,2016ApJ...824...96T,2016ApJ...823..110R}. EBs are seen as enhanced intensities between 30\% to 55\% above average brightness in the wings of the H$\alpha$ line profile, often present above the polarity inversion line \citep{2007A&A...473..279P,2008ApJ...684..736W,2016ApJ...823..110R}. 

EBs are also observed in other lines. \citet{2016ApJ...824...96T} observed EB like events in Mn~{\sc i} 2795 \AA, Mg~{\sc ii} h and k lines, Ni~{\sc ii} 1393.33 \AA, and 1335.30 \AA\ as enhancements in the wings rather than the core. EBs are observed in the Solar Dynamic Observatory's \citep[SDO,][]{2012SoPh..275....3P} Atmospheric Imaging Assembly \citep[AIA,][]{2012SoPh..275...17L} 1700 channel as small brightenings. Out of the ten events that \citet{2016ApJ...824...96T} identified as UV bursts \citep{2014Sci...346C.315P,2015ApJ...812...11V}, seven were along the magnetic inversion line, and three were co--spatial with EBs. \cite{2000ApJ...544L.157Q}, show that there is a significant correlation with EBs in the H$\alpha$ wings at $\pm$ 1.3 \AA\ and the UV continuum at 1600 \AA. \cite{2006ApJ...643.1325F} and \cite{2007A&A...473..279P}, noted the presence of EBs in the Ca~{\sc ii} 8542 \AA\ lines. Spectropolarimetric observations done by \citet{2007A&A...473..279P}, show that EBs are formed when opposite polarities merge giving rise to a cancellation of magnetic flux. As this cancellation occurs, plasma is heated and accelerated deep in the atmosphere and this is seen as a double--shaped hump in IRIS's Si~{\sc iv}, C~{\sc ii} and Mg~{\sc ii} lines. The total energies needed to produce EBs are estimated to be in the range of 10$^{27}$ to 10$^{28}$ ergs \citep[and references therein]{2002ApJ...575..506G}, however in the IRIS observations the energy needed to drive the UV bursts is of the order of 10$^{29}$ ergs \citep{2014Sci...346C.315P}. \citet{2016A&A...592A.100R} studied Ellerman Bomb-like brightenings in the quiet Sun (QS) and suggested that these EBs can only be identified at the telescope's diffraction limit of $\lambda$/D=0.14$''$ at 6563 \AA\ in SST data at a much lower intensity change, thus relaxing the \~50\% above average intensity requirement usually used to define EBs. Such QS observations of EBs were also reported by \cite{2017ApJ...845...16N}.  \cite{2017ApJ...845..100R} report micro-flaring events that are in some cases similar to the classical definition of EBs and discussed the need for redefining EBs, based on signatures depending only on observations.


\subsection{Magnetic concentrations and pseudo-EBs}
\cite{2013ApJ...774...32V} classify an EB when the mean intensity enhancements are between 30\% to 55\% in the H$\alpha$ line wings, as compared to the average background line profiles. They further show that the bright grains, that are found simultaneously in the Ca~{\sc ii} H, and the G-band images are bright network points. Such network bright points are driven by strong magnetic field concentrations \citep{1969SoPh....9..347S,1971IAUS...43..329V,1973SoPh...28...61H,1987SoPh..112..295M,2005ApJ...635..659H}. 

\cite{1976SoPh...50..269S}, suggest that these magnetic concentrations (MCs) are bright in the continuum of hot--wall radiation. \cite{2004A&A...428..613B} and \cite{2005A&A...435..327R} indicate that the MCs rapidly evolve with complex morphologies. However, the MCs are found in the dark intergranular lanes and are only observed at a sub-arcsecond resolution \citep{1996ApJ...463..797T}. MCs are further observed in Mn~{\sc i} \citep{1987ApJ...314..808L}, line wings of H$\alpha$ \citep{2006A&A...449.1209L} and the G-band \citep{2006A&A...452L..15L}. They are less sharp in the Ca~{\sc ii} H. In the DOT movies, \cite{2013JPhCS.440a2007R} reports that MCs appear in the blue--wing of H$\alpha$, suggesting down flows. On comparing the signatures in H$\alpha$ and Na~{\sc i} D, they see MC shocks accompanied with blue--wing enhancements in H$\alpha$.

Furthermore, \cite{2013JPhCS.440a2007R} suggested that the mean intensity change in the wings of the H$\alpha$ line has to be at least 50\% with respect to the average background line profile, and all EBs fainter than this should be considered as pseudo--EBs irrespective of the formation mechanisms. Such a definition suggests that the 3500$+$ EBs studied by \cite{2013SoPh..283..307N} are pseudo--EBs. Additionally, \citet{2015ApJ...812...11V} suggested that the false-positives by \cite{2013SoPh..283..307N}, are because the regions studied were close to a decaying sunspot rather than an emerging sunspot. The most important difference between a MC and an EB is that EBs are related to reconnection.

\subsection{Quiet Sun EBs (QSEBs)} 

QSEBs have a similar topology to EBs, such as a bright flame and lifetimes of a few minutes. \citet{2016A&A...592A.100R} observed these events in the H$\alpha$, Fe I 6173 \AA, and Ca~{\sc ii} 8542 \AA, wavelengths in combination with IRIS and AIA/SDO. They found the EB intensities significantly lower than the active region EBs. However, the authors suggest that these EBs are also consequences of reconnection. Moreover, they also suggested that QSEBs are detected only when the data is of a high quality. Such data can be acquired from the Swedish 1-m Solar telescope (SST, \cite{2003SPIE.4853..341S}) and is enhanced with the support of the adaptive optic system and image reconstruction techniques such as Multi-Object Multi-Frame Blind Deconvolution (MOMFBD, \cite{2005SoPh..228..191V}). \citet{2016A&A...592A.100R} identify 24 QSEBs in a 4 Jul 2013 09:20 UT dataset and a further 21 QSEBs in a 4 Jul 2013 10:13 UT dataset. Furthermore, they describe QSEBs to have lengths between 150 to 360 km and widths $\approx$ 170 km. QSEBS are observed in positions H$\alpha$ $\pm$ 1.3 \AA\ and last for a few minutes. QSEBs tend to have a predominantly bipolar topology, where after reconnection, both polarities seem to diminish. The intensity enhancement was below 40\% in the H$\alpha$ line wing (in relation to the reference spectrum). Also, they concluded that these QSEBs are not observed in Ca~{\sc ii} as they couldn't find significant evidence. In addition, \citet{2017ApJ...845...16N}, show the presence of QSEBs in their dataset.

\subsection{Layout}
 
We show certain cases of enhancements in the wings of H$\alpha$, in the range of 10\%--20\% (above the QS average intensity) associated with the interaction of opposite polarities observed in Fe~{\sc i}. Most of these events are also observed in Ca~{\sc ii} line wings with some events showing core enhancement. Such Ca~{\sc ii} line wings enhancements were not reported with QSEBs before. We use light curves in H$\alpha$, Ca~{\sc ii}, Stokes-$V$ and Stokes-$I$ from Fe~{\sc i}, to investigate the evolution of QSEBs. Using a time series of magnetised photospheric models produced by the MURaM code \citep{2005A&A...429..335V}, we further analyse the character of plasma motions in intergranular magnetic field concentrations and in particular the formation height as seen in H$\alpha$ and Ca II 8542 \AA. Such an approach provides a comprehensive understanding of the source of heating associated with these QSEBs. 

 
\begin{figure*}
\centering
\includegraphics[width=1.0\textwidth, height=0.40 \textheight]{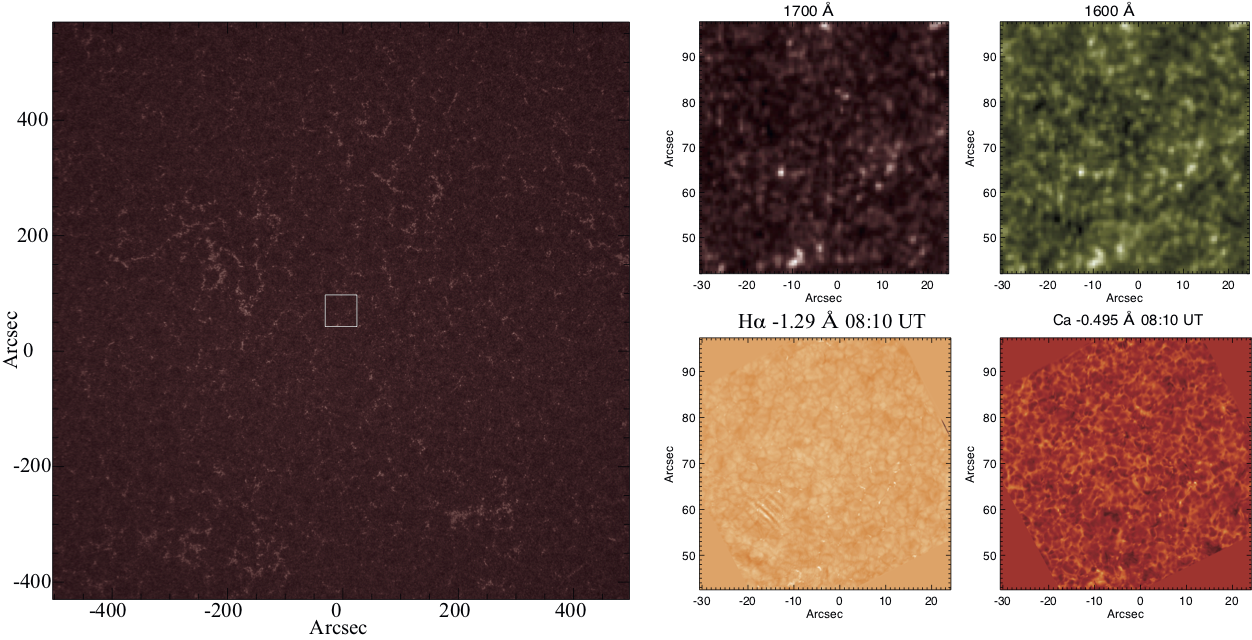}
\caption{The quiet Sun region observed using CRISP on SST. The left panel shows the location of the QS SST FOV overplotted against the AIA 1700 \AA\, channel. The panels on the right panel, show the zoomed-in view of the SDO AIA 1600 \AA\, and 1700 \AA\, channels, with corresponding H$\alpha$ (6563 \AA), Ca~{\sc ii} (8542 \AA) images obtained from SST. Crosses ("X") represent the locations of 10 selected events and A corresponds to a unipolar event.}
\label{intro}
\end{figure*}

\begin{figure*}
\centering
\includegraphics[width=1.0\textwidth, height=0.40 \textheight]{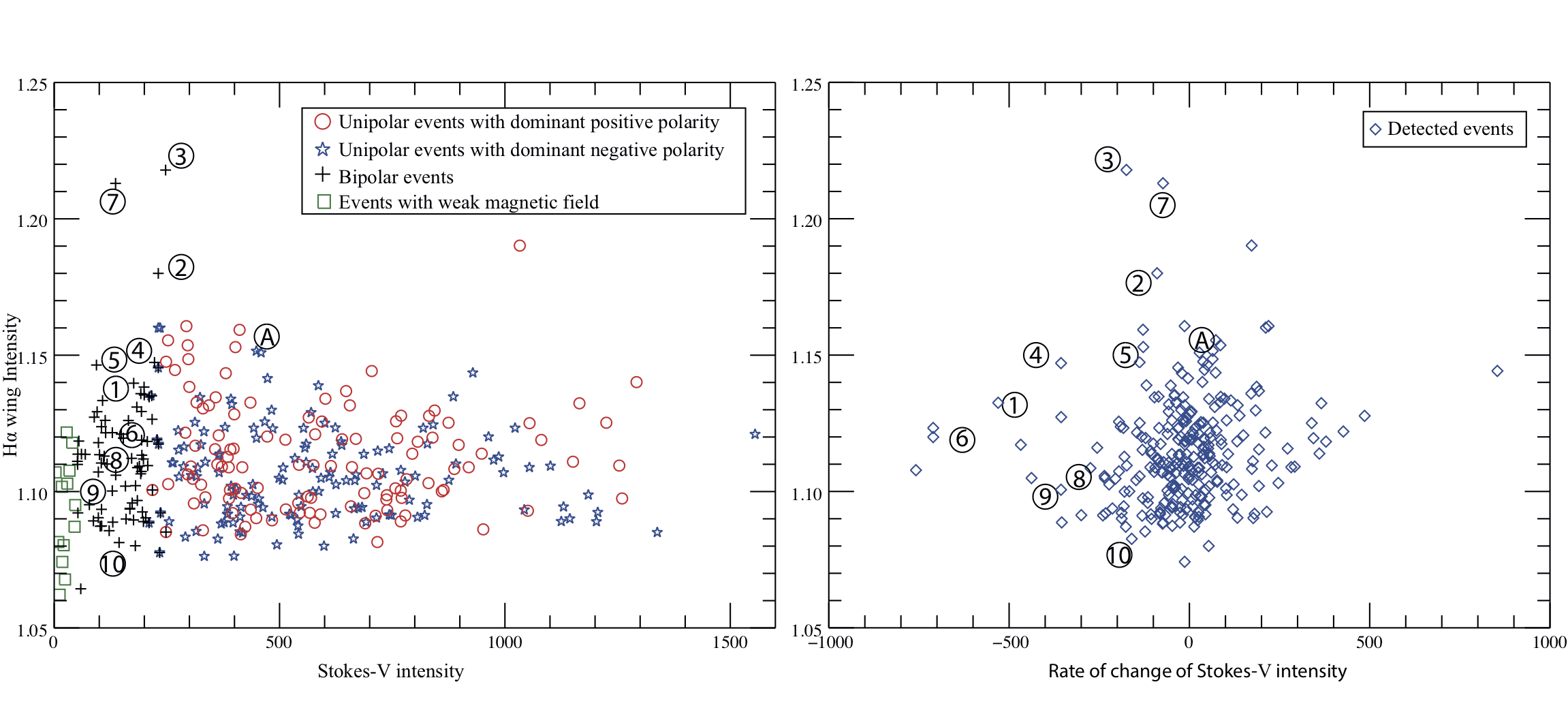}
\caption{Scatterplot of detected events representing: in the left panel a relation between fractional change in the H$\alpha$ wing intensity and the Stokes-$V$ \textbf{amplitude}, and in the right panel the relationship between the fractional change in the H$\alpha$ wing intensity and the rate of change of the Stokes-$V$ signal. In the left panel the unipolar regions are represented by red-circles and blue-stars. Bipolar regions are represented by black-plusses. The events with less than 50 units of Stokes-$V$ signal (weak events) are represented by green-squares. The manually selected regions are represented by numbers 1--10, and "A" represents a sample unipolar event. In the right panel, all the events are represented by blue diamonds with the selected events represented by 1--10 with a sample unipolar region represented as "A" (see text for details).}
\label{scatterplot}
\end{figure*}


\begin{figure*}
\centering
\includegraphics[width=0.99 \textwidth, height=0.63 \textheight]{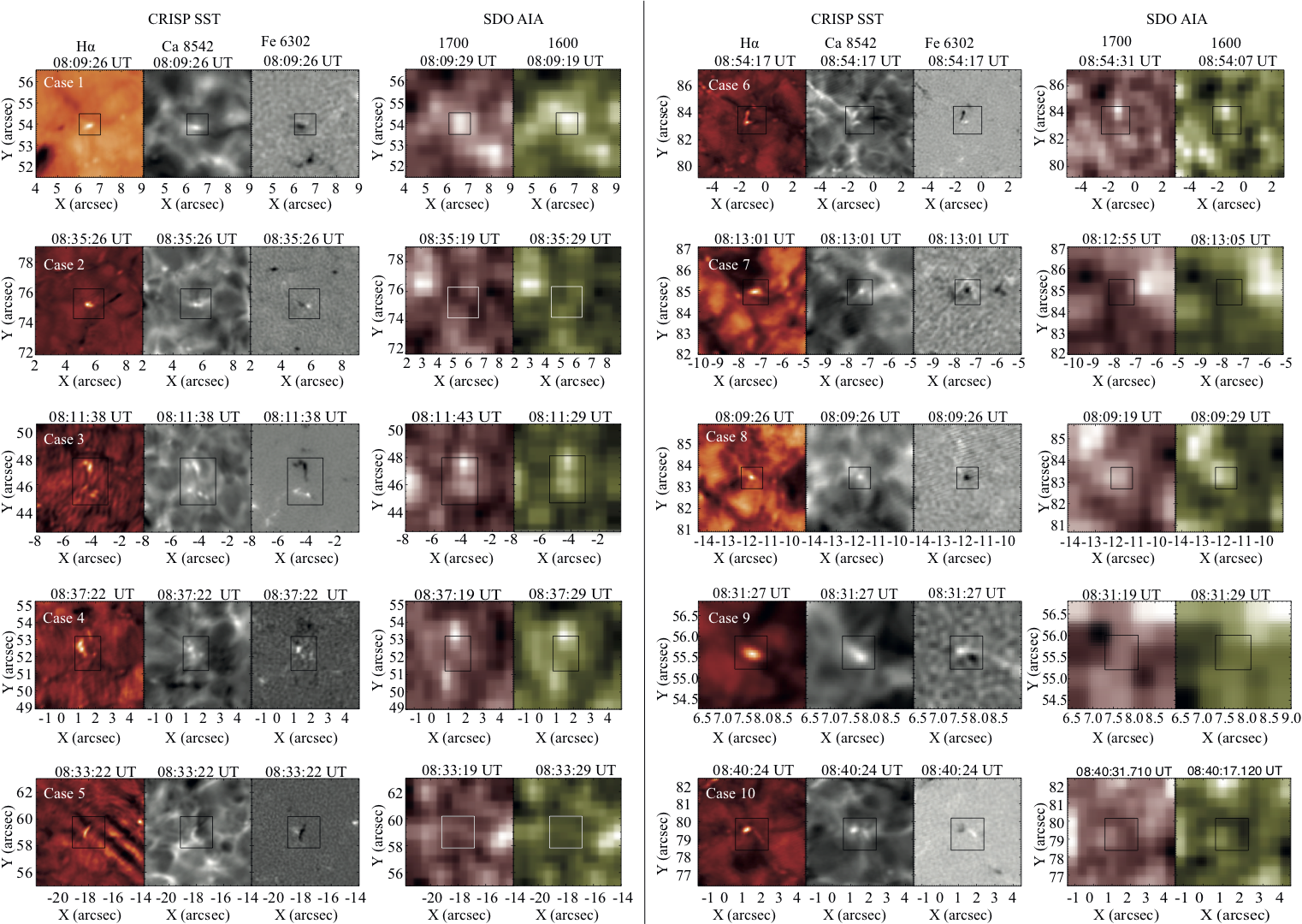}
\caption{Snapshots of 10 possible QSEBs. The panels show the event in H$\alpha$ (first column) wing position -1.29 \AA, Ca~{\sc ii} 8542 \AA\ wing position -0.495  \AA\ (second column) and Fe~{\sc i} 6302 \AA\ Stokes-$V$ (third column). The AIA data from 1600 \AA\ and 1700 \AA\ is shown in the fourth and fifth columns respectively. The boxes overplotted on the images show the location of the QSEBs.} 
\label{QSEB_snapshots}
\end{figure*}

\vspace{-0.6cm}
\section{Observations} \label{obs}
We investigate a quiet Sun disk center dataset, taken between 08:07:24 -- 09:05:46 UT on 21 June 2012 using CRisp spectro-polarimeter \citep[CRISP,][]{2008ApJ...689L..69S} on the Swedish 1--m Solar Telescope \citep[SST,][]{2003SPIE.4853..341S}. We use QS observations in H$\alpha$ (6563 \AA\,), Ca~{\sc ii} (8542 \AA\,), and Fe~{\sc i} (6302 \AA\,). The FOV was centred in the QS at [-3.1$''$,69.9$''$]. Fig~\ref{intro} shows the location of the FOV against AIA 1700 \AA\, channels. The panels show the zoomed-in view of SDO AIA 1600 \AA\, and 1700 \AA\, channels, with corresponding H$\alpha$ (6563 \AA), Ca~{\sc ii} (8542 \AA) images. Crosses ("X") represent the locations of 10 selected events and A corresponds to a unipolar event. Multi--Object Multi--Frame Blind Deconvolution (MOMFBD) data reduction was performed using the method by \citet{2005SoPh..228..191V}. In H$\alpha$, we observed at 10 line positions corresponding to $\pm$ 1.29 \AA, $\pm$ 1.03 \AA, $\pm$ 0.774 \AA, $\pm$ 0.516 \AA, $\pm$ 0.258 \AA\ from the line center at 6563 \AA\ (corresponding to Doppler velocities of $\pm$ 59 km s$^{-1}$,   47 km s$^{-1}$, 35 km s$^{-1}$, 23 km s$^{-1}$, 12  km s$^{-1}$. In Ca {\sc ii} (8542 \AA) we observe at $\pm$ 0.495  \AA, $\pm$ 0.440 \AA,  $\pm$ 0.384 \AA,  $\pm$ 0.330 \AA, $\pm$ 0.275 \AA,  $\pm$0.219 \AA,  $\pm$0.165 \AA, $\pm$ 0.110 \AA,   $\pm$0.054 \AA\ with respect to the line center at 8542 \AA. In Fe~{\sc i} (6302 \AA) we obtained spectro--polarimetric observations only at one position at about -40 m\AA\ from the line core. The cadence of this dataset is 8 s.

\begin{figure*}
\centering
\includegraphics[width=1.00 \textwidth, height=0.93 \textheight]{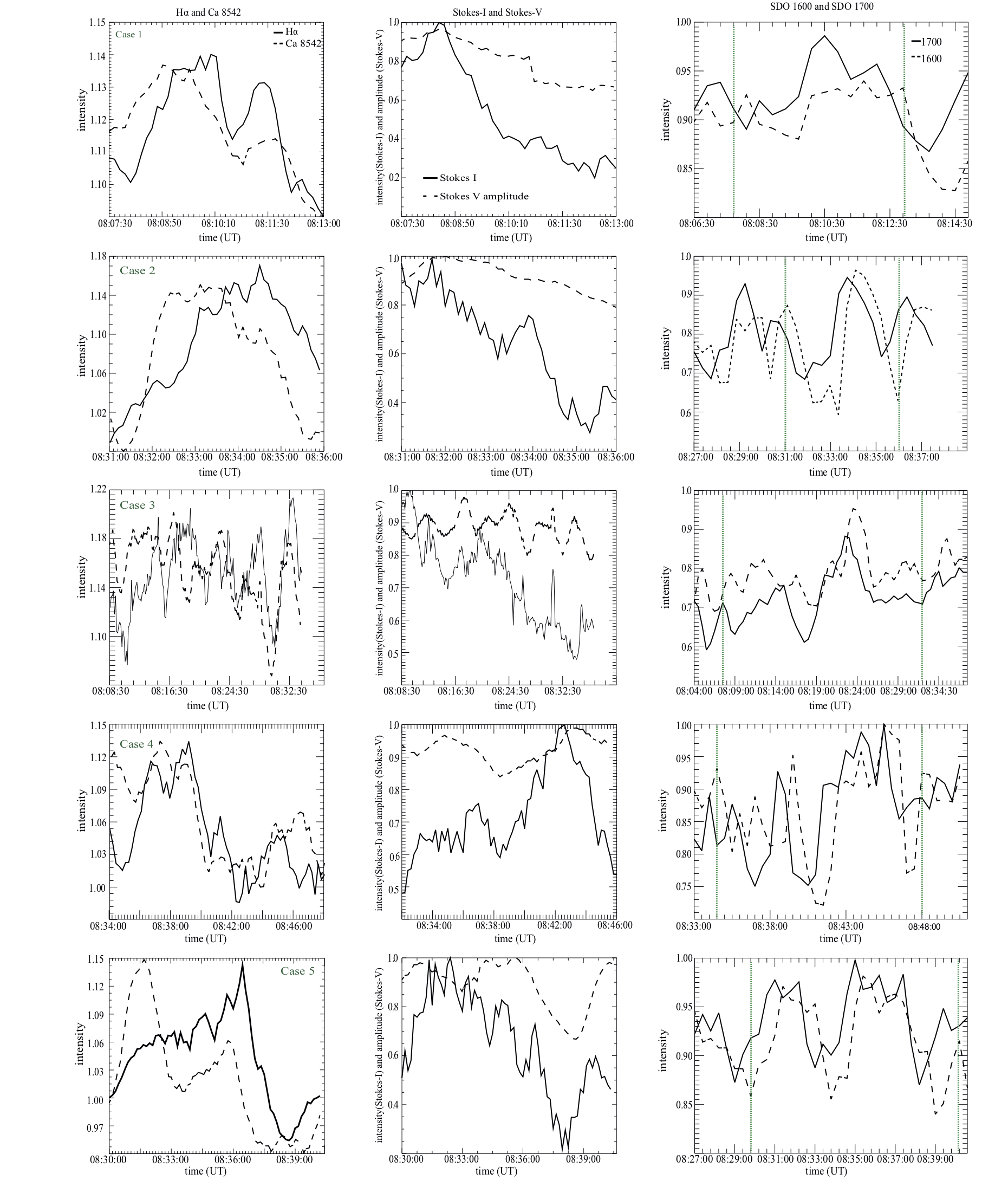}
\caption{Light curves for cases 1--5. Column 1: Light curves in H$\alpha$ (solid-black) wings at $\pm$ 1.29 \AA\ and Ca~{\sc ii} 8542 \AA\ (black dashed) wings at $\pm$ 0.495  \AA , Column 2:  Light curves representing amplitude of in Fe~{\sc i} 6302 \AA\ Stokes-$V$ (black dashed)  and Fe~{\sc i} 6302 \AA\ Stokes-$I$ (solid black). Column 3: Light curves in 1600 \AA\ (solid black)  and 1700 \AA\ (black dashed) channels obtained from SDO-AIA. The vertical green dotted lines overplotted on the light curves obtained in the 1600 and 1700 channel represent the start time and end time of the event as observed in H$\alpha$ (solid-black) and Ca~{\sc ii.}} 
\label{EB_lightcurves_1}
\end{figure*}

\begin{figure*}
\centering
\includegraphics[width=1.00 \textwidth, height=0.93 \textheight]{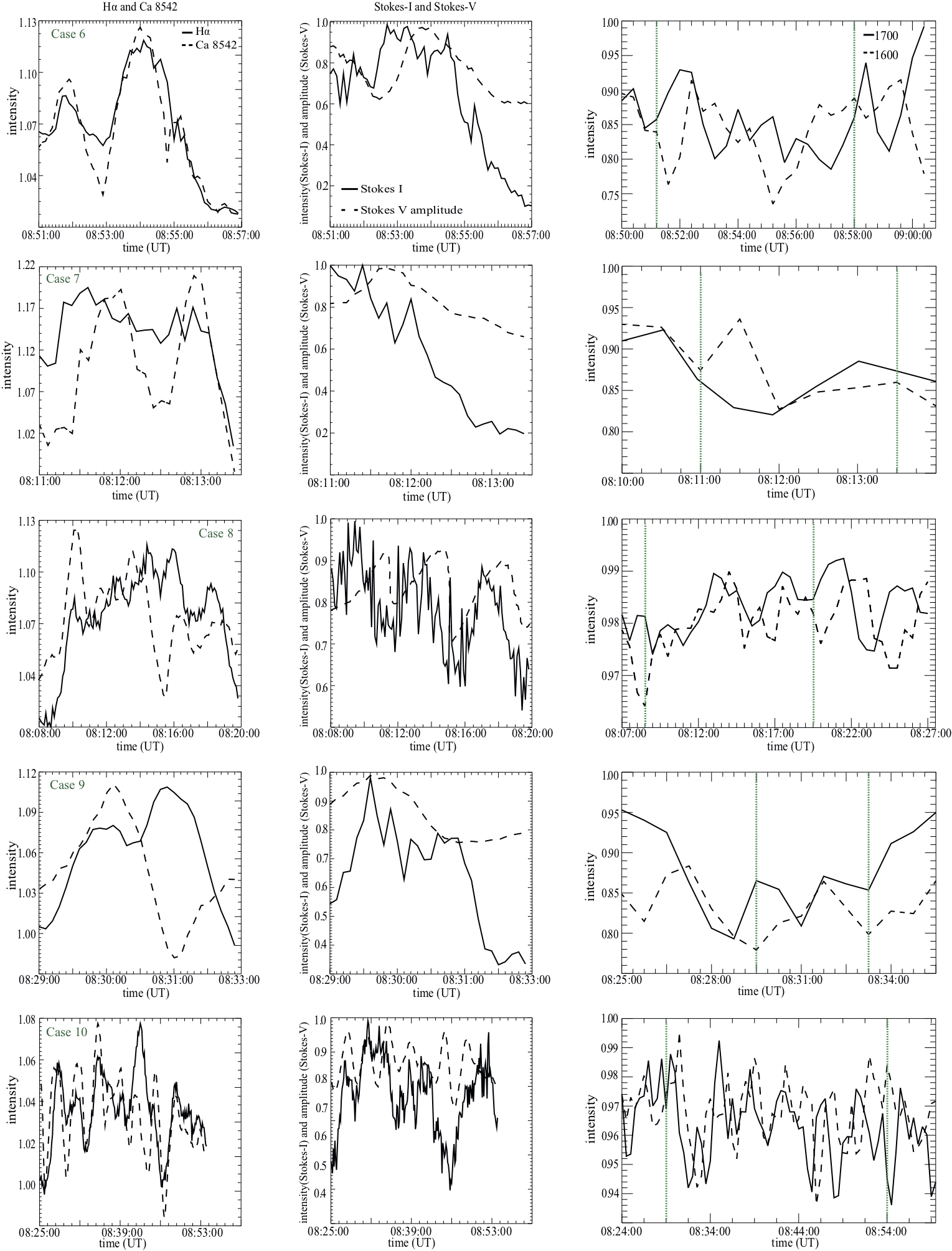}
\caption{Light curves for cases 6--10. Column 1: Light curves in H$\alpha$ (solid-black) wings at $\pm$ 1.29 \AA\ and Ca~{\sc ii} 8542 \AA\ (black dashed) wings at $\pm$ 0.495  \AA , Column 2:  Light curves representing amplitude of in Fe~{\sc i} 6302 \AA\ Stokes-$V$ (black dashed)  and Fe~{\sc i} 6302 \AA\ Stokes-$I$ (solid black). Column 3: Light curves in 1600 \AA\ (solid black)  and 1700 \AA\ (black dashed) channels obtained from SDO-AIA. The vertical green dotted lines overplotted on the light curves obtained in the 1600 and 1700 channel represent the start time and end time of the event as observed in H$\alpha$ (solid-black) and Ca~{\sc ii}.}
\label{EB_lightcurves_2}
\end{figure*}
 
\section{Detection method} \label{method}
We use the EB detection automation code $''$EBDATA$''$ by \citet{2016ApJ...823..110R}. The algorithm can detect and track magnetic concentrations. The algorithm relies on parameters of the magnetic concentrations such as intensity threshold, area, size and lifetime. The detection code also compares changes in intensity of the events with respect to surrounding intensity changes. In order to allow detections in the QS regions with low intensity changes (as compared to the background), we made minor changes to the algorithm. We define an event as a QSEB when the intensity contrasts, calculated in the H$\alpha$ wings, for at least one pixel is 10\% more than the background average intensity. QSEBs evolve in time. This evolution is in terms of lateral motion as well as growth in size. The intensity of the grown area has to be greater than 10\% of the background average intensity, in both wings of the H$\alpha$ at $\pm$1.29 \AA. The area of the QSEB, when it is fully grown has to be greater than 2 pixels (132 km). The line core in H$\alpha$ must remain unchanged (no more than 1\% increase to account for variability as per \cite{2016ApJ...823..110R}). We detected 334 events, which are summarised in Fig~\ref{scatterplot}. The left panel of Fig~\ref{scatterplot} shows a relation between the fractional change in the H$\alpha$ wing intensity and the apparent flux represented by the Stokes-$V$ amplitude. The right hand panel shows a relation between the fractional change in the H$\alpha$ wing intensity and the rate of change of Stokes-$V$ amplitude. Here, the maximum intensity is given as the maximum value in the wings of the detected pixels relative to the FOV average. The rate of change of Stokes-$V$ amplitude is computed from change of Stokes-$V$ amplitude throughout the lifetime of the events. Most of the events identified by the routine were unipolar magnetic concentrations, shown by red circles and blue stars in the left panel of Fig~\ref{scatterplot}. The rest were bipolar magnetic concentrations with possible EB-like wing enhancements. These are represented by black-crosses in the left panel of Fig~\ref{scatterplot}. The events with less than 50 units of Stokes-$V$ signal are termed as weak events and are represented by green squares in the left panel of Fig~\ref{scatterplot}. In addition, the automated procedure also detected some long lasting events with a strong unipolar field, which would lie on the right hand side of the left panel of Fig~\ref{scatterplot} between Stokes-$V$ amplitude of "1000-1500". This unipolar events would correspond to a very low change in Stokes-$V$ signal, and would lie near the "0" mark in the right hand panel. The right hand panel of Fig~\ref{scatterplot}, shows events, which showed flux cancellation on the left hand side, with negative flux signs. The events that showed emergence of flux are represented on the right-hand side of the plot, these show positive flux. All the selected events show flux cancellation.

We manually selected 10 events from the detected events shown in Fig~\ref{scatterplot} satisfying properties of active region EBs, where magnetic flux cancellations are accompanied with wing enhancements in H$\alpha$ and Ca~{\sc ii} 8542 \AA. The events shown here are further selected by manual detection, which focused on 1.) interaction of QSEBs in Fe~{\sc i} 6302 \AA\ Stokes-$V$ wavelength, 2.) sudden intensity enhancements in the H$\alpha$ wing positions and 3.) sudden intensity enhancements in the Ca {\sc ii} 8542 \AA\ wing positions. These intensity enhancements are smaller compared to regular active region EBs. After detecting events using the code, we manually checked whether they were formed above the interacting opposite polarity regions. Fig~\ref{scatterplot} shows scatter plots highlighting  the comparison between properties of all detections and selected events. These events are labelled by the numbers 1--10 in Fig~\ref{scatterplot}. For the selected events we present the snapshots of the H$\alpha$ wing position at -1.29 \AA, Ca~{\sc ii} 8542 \AA\ at -0.495 \AA, and Fe~{\sc i} 6302 \AA\ Stokes-$V$ as well as their appearance in the SDO AIA's 1600 and 1700 \AA\ wavelengths in Fig~\ref{QSEB_snapshots}. The white and black boxes are overplotted on the images. These boxes represent the region of interest, which are then used for further analysis. Three of the selected events have recurring intensity enhancements. such EB recurrence have also been seen in active regions \citep{2000ApJ...544L.157Q,2015ApJ...798...19N, 2016ApJ...823..110R}. 

\begin{figure*}
\centering
\includegraphics[width=0.80\textwidth, height=0.85 \textheight]{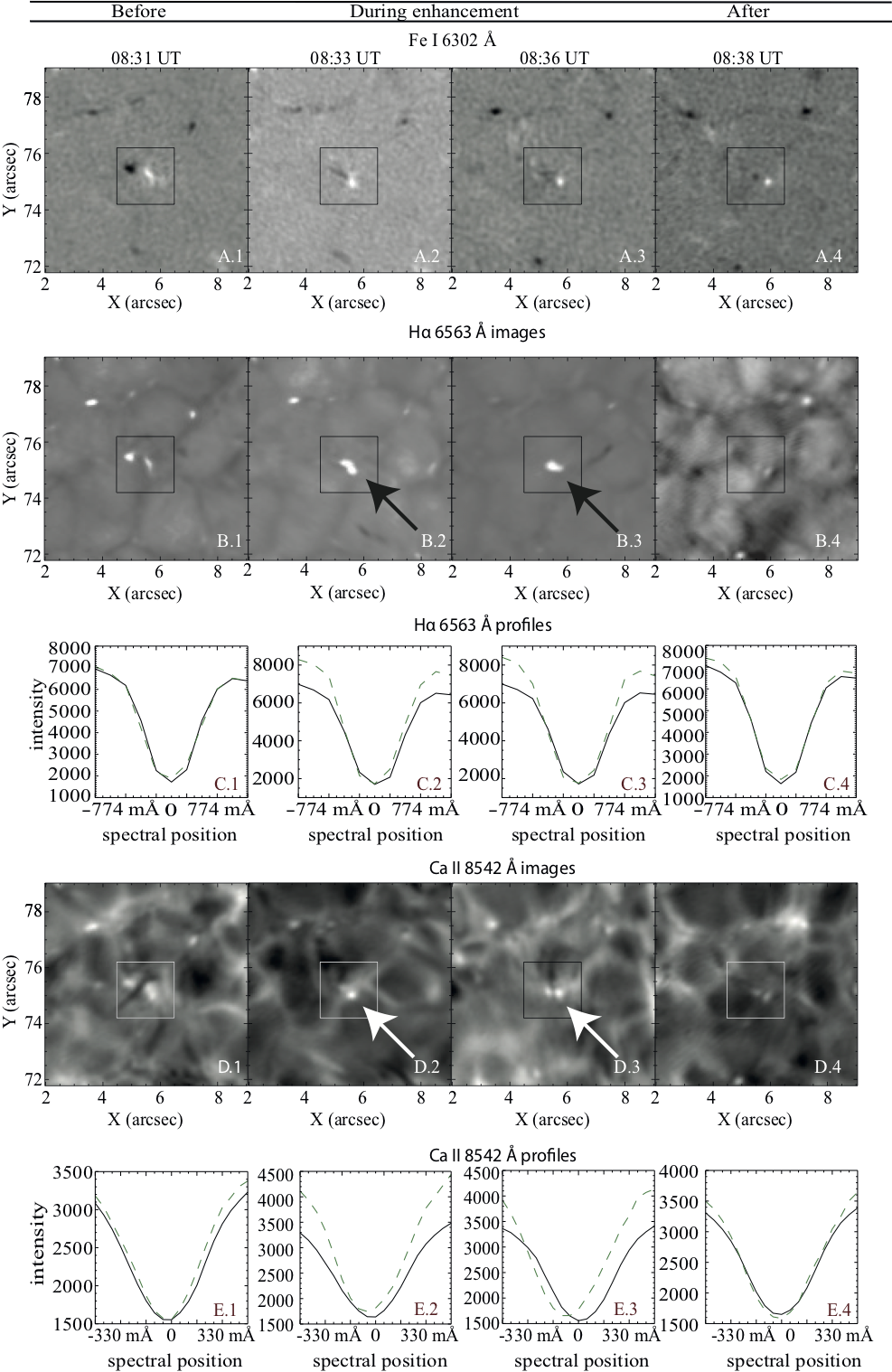}
\caption{Evolution of the EB represented as Case 2. The top panel A1--A2, are snapshots of interacting opposite polarities as seen in Fe~{\sc i} 6302 \AA. In the middle panel (B1--B2), we see an EB--like formation in the H$\alpha$ wings -1.29 \AA\ and Ca~{\sc ii} images at -0.495 \AA (D1--D2). The locations of intensity enhancements are indicated by arrows. The line--profiles in H$\alpha$ are shown in panels C1--C4 and Ca~{\sc ii} line--profiles are shown in E.1--E2. The green dashed lines represent the event and the averaged background line profile is represented by the solid black lines.}
\label{EB_case3}
\end{figure*}

\begin{figure*}
\centering
\includegraphics[width=0.85\textwidth, height=0.9 \textheight]{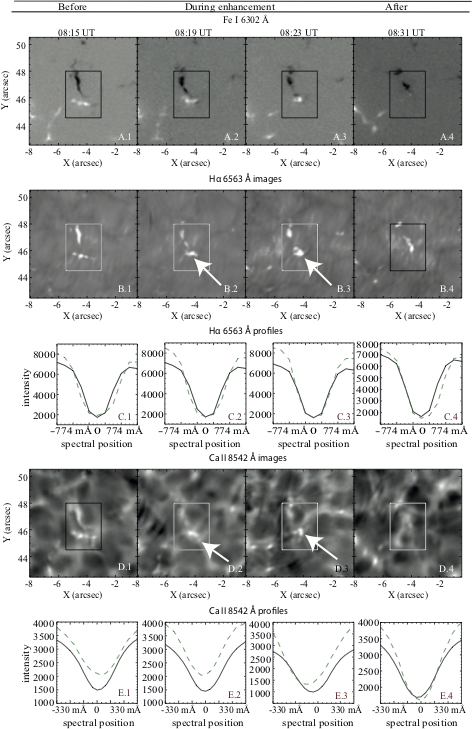}
\caption{Evolution of the EB represented as Case 3. The top panel A1--A2 are snapshots of interacting opposite polarities as seen in Fe~{\sc i} 6302 \AA. In the middle panel (B1--B2), we see an EB--like formation in the H$\alpha$ wings -1.29 \AA\ and Ca~{\sc ii} images at -0.495 \AA (D1--D2). The locations of intensity enhancements are indicated by arrows. The line--profiles in H$\alpha$ are shown in panels C1--C4 and Ca~{\sc ii} line--profiles are shown in E.1--E2. The green dashed lines represent the event and the averaged background line profile is represented by the solid black lines.} 
\label{EB_case8}
\end{figure*}

\section{Analysis and Results} \label{anal}

Fig~\ref{scatterplot} shows the manually selected events labelled 1--10. In the left hand side panel of Fig~\ref{scatterplot}, the selected events lie in the region of bipolar detections with H$\alpha$ wing intensities between 1.06--1.23 and a high reduction in the Stokes-$V$ intensity. A unipolar scenario event identified by the detection algorithm is labelled $''$A$''$ in the plot. This event is not considered as a QSEB. This event shows maximum intensity similar to some bipolar detections however, since the event does not evolve during its lifetime, the rate of change of Stokes-$V$ intensity  is nearly zero. The same goes with other events identified by the detection code. The manually selected events are bipolar MCs with a significant reduction in the Stokes-$V$ intensity. Thus the manually selected events stand out and have different properties from the other unipolar/bipolar events.

\subsection{Light curves}

In column 1 of Fig~\ref{EB_lightcurves_1} and Fig~\ref{EB_lightcurves_2}, we plot H$\alpha$ (solid black)  and Ca~{\sc ii} 8542 \AA\ (black dashed) light curves. The H$\alpha$ light curves are computed at $\pm$1.29 \AA, and the Ca~{\sc ii} 8542 \AA\ light curves are computed at $\pm$0.495 \AA. We plot light curves for the selected events by taking all pixels relating to the feature and summing the intensities in the blue and the red wing of the H$\alpha$ and Ca~{\sc ii} 8542 \AA\ lines. The output is then divided by average value. The light curves show a impulsivity corresponding to EB--like line wing enhancements. We see enhancements in the H$\alpha$ and Ca~{\sc ii} 8542 \AA\ wings. The minimum intensities of the light curves correspond to times where the H$\alpha$ intensity contrast came back to the averaged background intensity. The maximum intensities correspond to our events observed in the H$\alpha$ and Ca~{\sc ii} 8542 \AA\ wing. We plot light curves in Stokes-$V$ (solid black) and Stokes-$I$ (black dashed) signal at -40 m\AA\ from the Fe 6302 \AA\ line centre. The light curves are shown in column 2 (middle column). The light curves are plotted in black solid and dashed colours respectively in Fig~\ref{EB_lightcurves_1} and Fig~\ref{EB_lightcurves_2}. We see that as one polarity disappears, the net-flux also reduces, indicating a cancellation of the magnetic field. In the column 3 of Fig~\ref{EB_lightcurves_1} and Fig~\ref{EB_lightcurves_2} we plot light curves corresponding to the SDO-AIA channels 1600 \AA\ (solid black) and 1700 \AA\ (black dashed). These are plotted by averaging all pixels corresponding to the events and dividing by the maximum intensity. These light curves represent the AIA 1600 and 1700 channels before and after the events. The green dotted vertical lines show the locations of the events as observed in CRISP. 
 


From Fig~\ref{EB_lightcurves_1} and Fig~\ref{EB_lightcurves_2} we see that in the most cases (9 out of 10) the peak in intensity light curves occur in the wings of Ca~{\sc ii} 8542 \AA\ before H$\alpha$. The offsets at these positions are shown in Table~\ref{downflow}. We can calculate the velocity of the down flow of the intensity enhancement. The approximate heights between the formation of H$\alpha$ and Ca~{\sc ii} 8542 \AA\ is assumed to be 500km from models by \cite{2009ApJ...694L.128L} and \cite{2012ApJ...749..136L}. The propagation speed of the enhancements are approximately calculated using (distance/time) as 6 km s$^{-1}$. For the EBs showing recurring activity these velocities are calculated by using the difference between most prominent peaks in the light curves of H$\alpha$ and Ca~{\sc ii} 8542 \AA. The impulsive nature of the events is associated with a corresponding decrease in the Stokes signals, thus suggesting that the intensity enhancements correspond to the magnetic flux cancellation possibility mentioned in \cite{2002ApJ...575..506G}. On comparing the 1600 and 1700 \AA\, light curves with the H$\alpha$ and Ca~{\sc ii} 8542 \AA\ the intensity peaks observed in 1600 and 1700 \AA\, occur after the main intensity peaks observed in the H$\alpha$ and Ca~{\sc ii} 8542 \AA\ (see cases 1, 3, 4 and 5). In some cases there is no apparent signature in the SDO channels (see case 6, 7 and 9). There are some cases which show a brightening in SDO-AIA channels at the location of the event. This brightening lasts for only one SDO frame (cases 2, 8 and 10). It is not clear whether these brightenings corresponds to the event as it often appears after a delay. In addition, the light curves of cases 3, 8 and 10 in Fig~\ref{EB_lightcurves_1} and Fig~\ref{EB_lightcurves_2}, show multiple impulsive bursts. Such behaviour is analogous to EBs observed near a large source of magnetic flux. The light curves show that the QSEBs presented here form at an atmospheric level a few hundred kms above the photospheric continuum. This is shown by the different timings of the peak intensity between Ca~{\sc ii} 8542 \AA\ and H$\alpha$ light curves. 

\subsection{Categories}
By obtaining observations across multiple wavelengths, we can correlate the physics involved in events like EBs. In Fig~\ref{QSEB_snapshots}, we show snapshots of 10 cases. The snapshots are taken at the H$\alpha$ wing position -1.29 \AA, Ca~{\sc ii} 8542 \AA\ wing position -0.495 \AA, and Fe~{\sc i} 6302 \AA\ Stokes-$V$. The boxes represent the ROI. Based on the observations, events can be categorised into the following evolutionary characteristics: 1.) Single impulsive events involving reduction in Stokes-$V$ after the intensity peaks in H$\alpha$ and Ca~{\sc ii} 8542 \AA\ wings as in the light curves for cases 1, 2, 7 and 9. In case 9, the  Ca~{\sc ii} 8542 \AA\ and Fe~{\sc i} 6302 \AA\ Stokes-$V$ track each other more closely than H$\alpha$, which is a common observational effect of dynamics related to magnetic concentrations. However, this reduction in intensity is followed by interaction of opposite polarity magnetic concentrations that gives rise to impulsivity. 2.) Events associated with reduction in Stokes-$V$ signal where the two polarities keep on interacting (cases 3, 8, and 10). This reduction is observed with a repetitive impulsive nature in the H$\alpha$ and Ca~{\sc ii} 8542 \AA\ wings, during the time of the interactions. 
The presence of Ca~{\sc ii} 8542 \AA\ wing emissions in addition to H$\alpha$ emissions in all the cases suggest that such events are triggered in the lower chromosphere. The events studied show lifetimes of $\sim$800 s with intensity change of <10 \% in comparison to the average spectral lines. The velocity corresponding to the lateral motion of the selected events lies in the range of 0.3 km s$^{-1}$ and 2.4 km s$^{-1}$. This velocity is computed by $''$ EBDATA$''$ detection algorithm. This velocity range matches with EBs found near active regions \citep{1987SoPh..108..227Z}.

\begin{table}
 \caption{Calculation of plasma velocity.}
 \label{downflow}
 \begin{tabular}{lcc}
  \hline
  Case No & $\Delta$T & Velocity of plasma.  \\
   & s & (Km/s)\\
  \hline
  1 & 60 & 8.33\\
 2 & 120 & 4.15 \\
 3 & \textbf{420}  & \textbf{1.21}  \\
 4 & 80 & 6.25\\
 5 & 260 & 1.92\\
 6 & 10 & 50\\
 7 & \textbf{-30} & \textbf{+16.4}\\
 8 & \textbf{300} & \textbf{1.51}\\
 9 & 640 & 0.78 \\
 10 & \textbf{330} & \textbf{1.51}\\
  \hline
 \end{tabular}
\end{table}

\subsection{QSEB morphology}\label{strong_EB}

In Fig~\ref{EB_case3}, we show a small EB--like event. In the top-most row with panels A.1--A.4,  we see that the two polarities interact continuously in the Fe~{\sc i} Stokes-V evolution. Such interaction gives rise to an enhanced emission in the H$\alpha$ wing images taken at $\pm$ 1.29 \AA\ and Ca~{\sc ii} 8542 \AA\ line profile at $\pm$ 0.495 \AA. The panels B.1--B.4 and D.1--D.4 of Fig~\ref{EB_case3} show snapshots taken in the H$\alpha$ wing position -1.29 \AA\ and Ca~{\sc ii} 8542 \AA\ wing positions at -0.495 \AA. We see typical EB topologies in both H$\alpha$ and Ca~{\sc ii} 8542 \AA\ images (see arrows in Fig~\ref{EB_case3}). In panels C.1--C.4 and E1--E.4 of Fig.~\ref{EB_case3}, we show snapshots of the line--profiles with dashed green lines against background line profiles (the background line--profile is the average background across the FOV) shown in solid black lines. We see that there is a contrast change between 10\% -- 20\% while comparing to the average H$\alpha$ spectrum. In Ca~{\sc ii} 8542 \AA\ we see that such events have higher contrast changes from 20\% to 40\% as compared to the average Ca~{\sc ii} 8542 \AA\ spectrum. However, the line profile in Ca~{\sc ii} 8542 \AA\ is asymmetric, see panels E.3 of Fig~\ref{EB_case3}. 

In Fig~\ref{EB_case8}, we show an event which involves two magnetic concentrations interacting for $\sim$15 mins. The photospheric flux cancellation is followed by repetitive emissions in the H$\alpha$ and Ca~{\sc ii} wings. In the panels A.1--A.4, we see opposite polarities interacting in the Fe~{\sc i} 6302 \AA\ line core images, in the evolution. The interaction between negative and positive polarity causes the weaker polarity to be annihilated over the evolution (not shown here). Furthermore, the panels A.1--A.4, show evolution of the two polarities where the positive polarity is seen to diminish in size in panel A.4 as compared to A.1. This merging and interaction gives rise to multiple intensity peaks, seen in the H$\alpha$ wing position -1.29 \AA\ and Ca~{\sc ii} 8542 \AA\ -0.495 \AA\ images (see panels B.1--B.4 and D.1--D.4). Below both the H$\alpha$ and Ca~{\sc ii} 8542 \AA\ images, we show snapshots representing line--profiles with dashed green lines against solid black lines which represent the average background spectrum (see panels C.1--C.4 and E.1--E.4).

 
\subsection{A sample unipolar event} \label{uni}

\begin{figure}
\centering
\includegraphics[width=0.48\textwidth, height=0.68 \textheight]{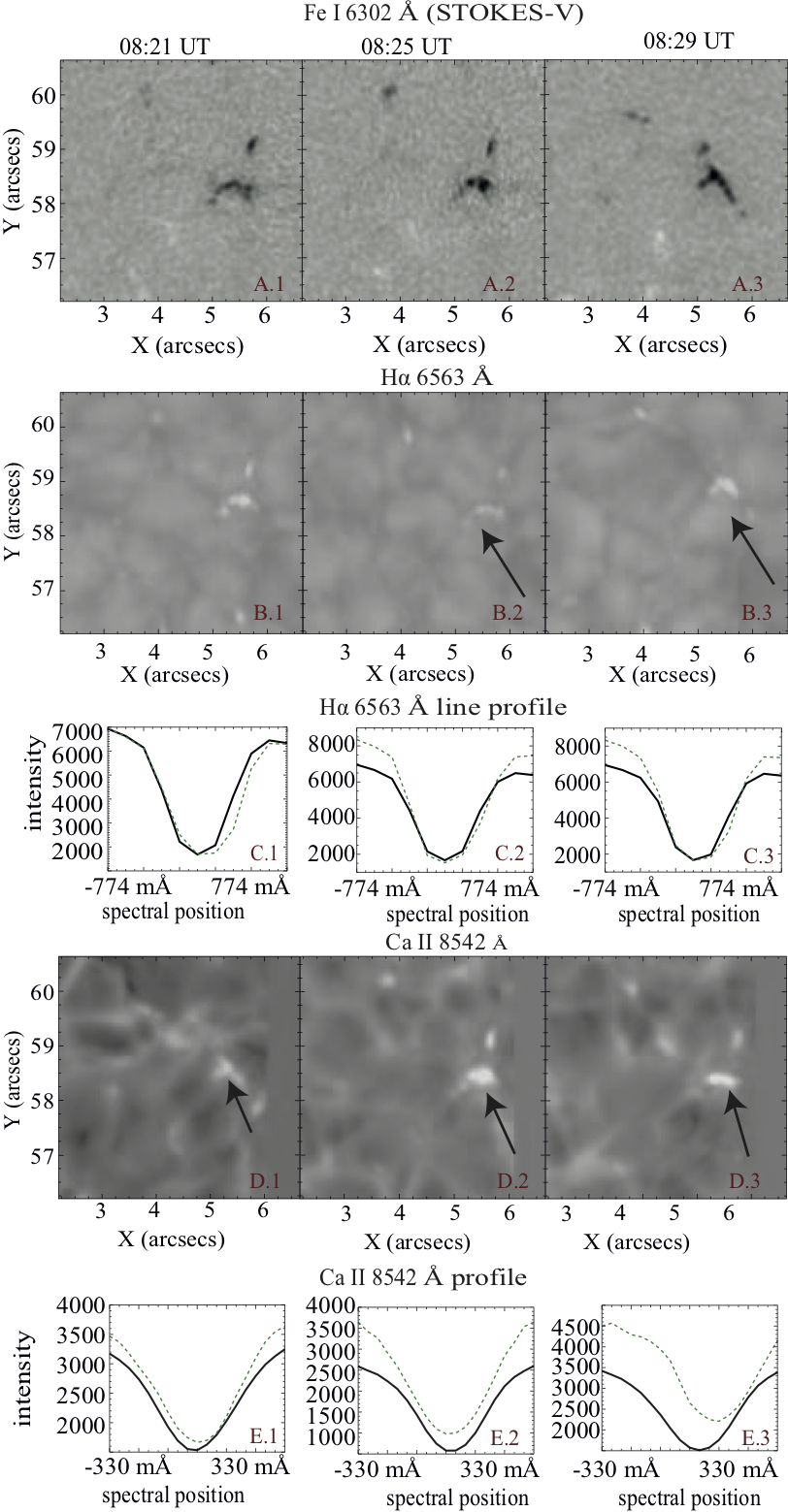}
\caption{Top panels shows a unipolar event with dominant negative polarity (indicated by "A" in Fig~\ref{scatterplot}). The top panel A1--A2, represents snapshots of interacting opposite polarities as seen in Fe~{\sc i} 6302 \AA. In the middle panel (B1--B2), we see a EB--like formation in the H$\alpha$ wings -1.29 \AA\ and Ca~{\sc ii} images at -0.495 \AA (D1--D2). The locations of intensity enhancements are indicated by arrows. The line--profiles in H$\alpha$ are shown in panels C1--C4 and Ca~{\sc ii} line--profiles are shown in panels E.1--E2. The green dashed lines represent the event and the averaged background line profile is represented by the solid black lines.}
\label{EB_case4}
\end{figure}

\begin{figure}
\centering
\includegraphics[width=0.45\textwidth, height=0.25 \textheight]{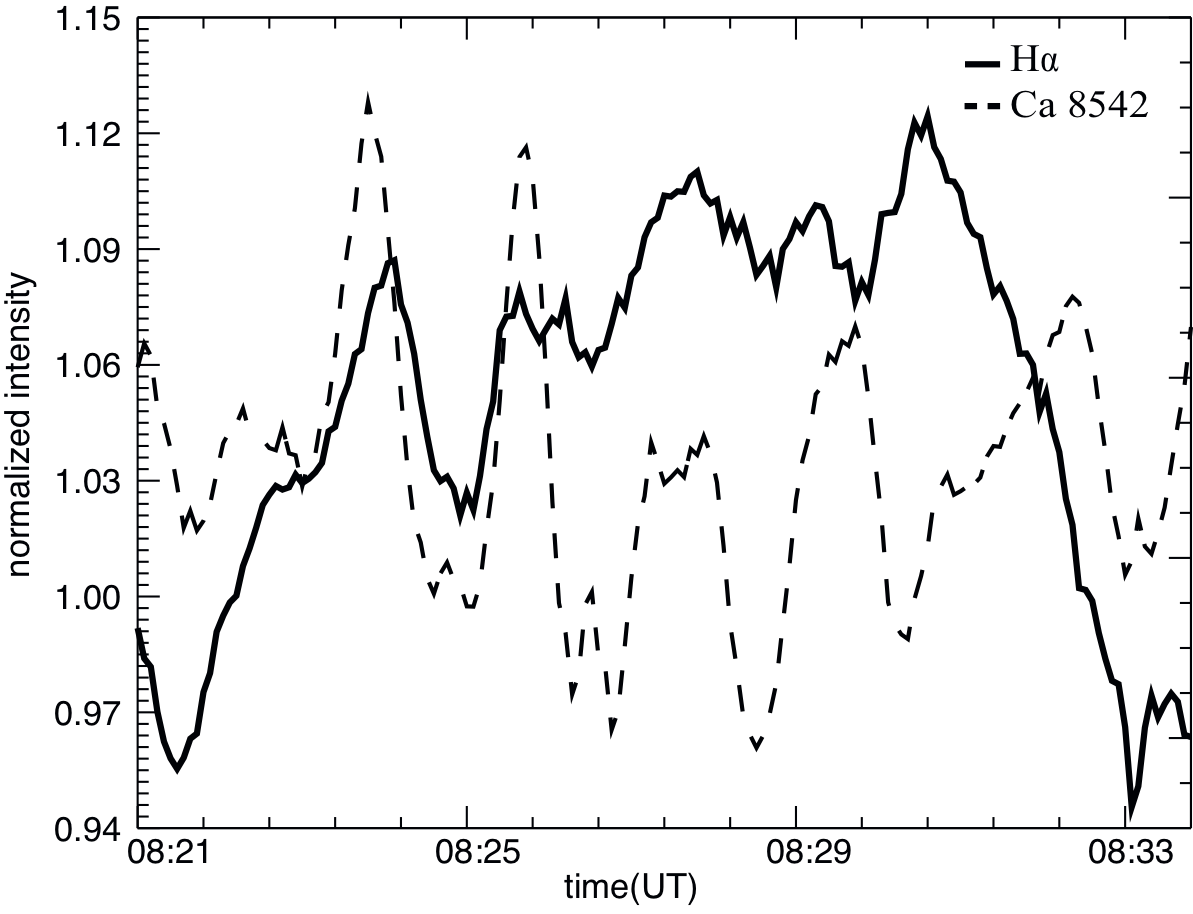}
\caption{Light curves for unipolar region shown in Fig~\ref{EB_case4}.  Light curves in H$\alpha$ (solid-black) wings at $\pm$ 1.29 \AA\ and Ca~{\sc ii} 8542 \AA\ (black dashed) wings at $\pm$ 0.495  \AA.}
\label{EB_case4_lightcurve}
\end{figure}

The EBDATA algorithm detected 334 events, out of which 10 were selected for detailed analysis. We discuss here the evolution of a unipolar event that was discarded as a $''$false positive$''$. The unipolar event is labeled $''$A$''$ in the Fig~\ref{scatterplot}. The snapshots of the evolution of this unipolar event are shown in panels A.1--A.3 of Fig~\ref{EB_case4} with larger negative polarity seen in the Fe~{\sc i} 6302 \AA\ Stokes-$V$. We see that the two unipolar flux regions interact with each other combining to form bigger negative polarity in size (sub panels A.1--A.3 of Fig~\ref{EB_case4}). In panels B.1--B.3, we show a series of the H$\alpha$ images. We see enhancements in the intensity at locations where the unipolar flux region combines. Such intensity enhancements are also seen in panels D.1--D.4. In both the H$\alpha$ and Ca~{\sc ii} images, we see EB-like wing enhancements. The H$\alpha$ and Ca~{\sc ii} 8542 \AA\ line profiles show a similar behaviour of emission as compared to previous examples (see the snapshots of line profiles seen in panels C.1--C.3 and E.1--E.3 respectively). However, Ca~{\sc ii} 8542 \AA\, line profile shows a strong blue--shifted line profile with core enhancements. Such events could be due to shearing reconnection, low-resolution imaging fails to spot the opposite polarity, or they could be driven by braided reconnection. Furthermore, Fig~\ref{EB_case4_lightcurve} shows no clear relation between the H$\alpha$ (solid black) and Ca~{\sc ii} 8542 \AA\ (black dashed) light curves. The examples here show impulsivity observed in H$\alpha$, which may or may not be related to the QSEBs. Hence we have ignored such detections.

The observational diagnosis indicates that in disk-centre viewing along the radial direction, only the top of an EB is seen, which shields what lies underneath as noted in simulations by \citet{2017A&A...601A.122D}. Thus there is an absence of flame-like topology here. Also, in comparison with the 1600 and 1700 \AA\ channels, we see no particular correspondence with the EB signatures observed in H$\alpha$ and Ca~{\sc ii} 8542 \AA. However, we do note that in some cases there is some brightening that occurs after the initial EB brightening that could be related to these events. A possible explanation for the lack of UV enhancement could be due to the lower spatial resolution of the SDO AIA instrument, or the lower height at which the UV continua form. It is analogous to reports by \cite{2016A&A...592A.100R}. \cite{2017A&A...601A.122D} conclude that the strongest brightening corresponds to a significant temperature and density increase that occurs at the site of the cancellation of two magnetic features of opposite polarities. Furthermore, the authors also highlight that unipolar regions are also strong EB candidates when accompanied by flux cancellation. This highlights that many detected unipolar regions could be an EB candidate. \citet{2002ApJ...575..506G} suggest that flux cancellation is possible in unipolar regions by shearing reconnection. Furthermore, \cite{2017ApJ...839...22H} using BiFROST simulations suggest a weak brightening in Si IV associated with EBs.

\begin{figure*}
\centering
\includegraphics[width=0.7\textwidth, height=0.2 \textheight]{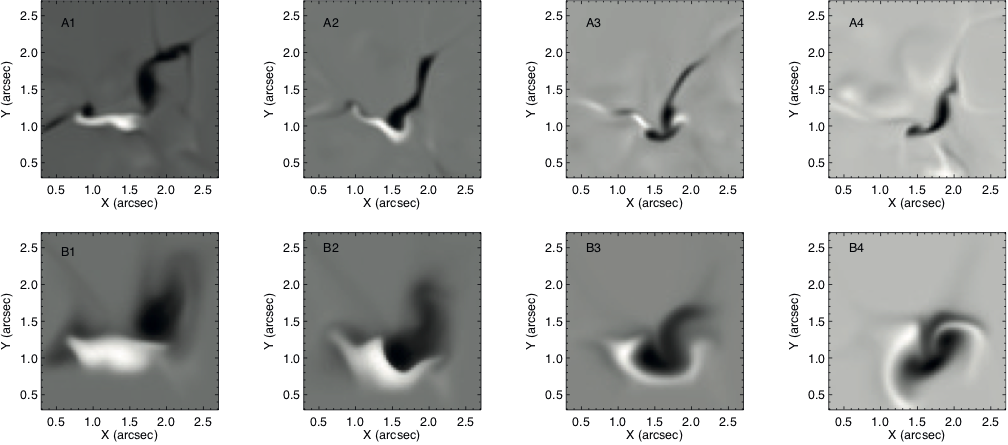}
\caption{Evolution of magnetic field concentrations in the MURaM simulations. The time-stamps are separated by 50s. The top row with panels A1--A4 shows interacting opposite polarity magnetic concentrations at the photospheric levels. The bottom row with panels B1--B2, show the responses of the interactions to magnetic concentrations at $\sim$430 km above the photosphere.}
\label{muram_fields}
\end{figure*} 

\begin{figure}
\centering
\includegraphics[width=0.46\textwidth, height=0.18 \textheight]{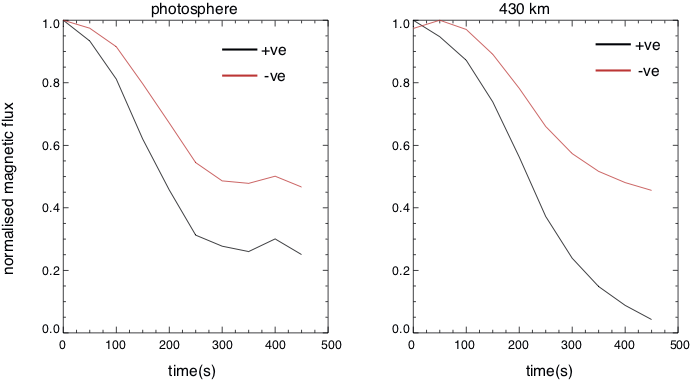}
\caption{The rate of change in magnetic flux at Fe~{\sc i}  6302 \AA\ as simulated in the MURaM simulations for the regions represented in Fig~\ref{muram_fields}. The left panel shows the rate of change in magnetic flux at the photospheric continuum level and the right panel shows rate of change in magnetic flux at 430 km above the photosphere.}
\label{muram_fe}
\end{figure}

\begin{figure*}
\centering
\includegraphics[width=0.7\textwidth, height=0.25 \textheight]{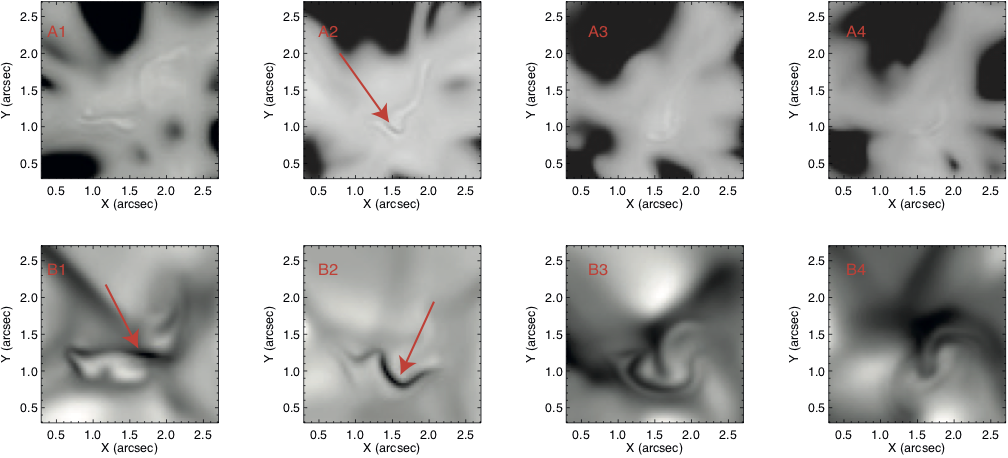}
\caption{Evolution of temperature, corresponding to panels in Fig~\ref{muram_fields} from the MURaM simulations. The time-stamps are separated by 50s. The top row with panels A1--A4, shows temperature change at the photospheric level. The bottom row with panels B1--B2, shows the temperature change at $\sim$430 km above the photosphere.}
\label{muram_temp}
\end{figure*} 

\section{MURAM simulations} \label{MURaM}

The aim of this simulation is to understand the formation mechanisms related to QSEBs. They complement the observations as the simulations performed in MURaM \citep{2005A&A...429..335V} tell us where these events are formed. The MURaM code is used to perform simulations of the interaction of the magnetic field concentrations in the solar photosphere. This particular setup is for a QS region. The numerical setup for these set of observations is similar to the one described in \cite{2013ApJ...779..125N}. The spatial resolution of the box is 25 km $\times$ 14 km $ \times$ 25 km. The temporal resolution of the simulation is 50 s. A positive-negative $''$checkerboard$''$ vertically-directed magnetic field, with the unsigned strength of $200~\mathrm{G}$ is added to a well-developed non-magnetic photospheric convection snapshot. Then the computational domain is set to evolve for a small number (2-5) of granular lifetimes. During the evolutionary period, most of the magnetic field cancels out, leaving some substantial magnetic field concentrations of opposite polarities in the intergranular lanes of the simulated photospheric granulation. These magnetic field concentrations move along the intergranular lanes occasionally coming in proximity to each other and reconnecting. 

Fig~\ref{muram_fields}, shows one such evolution for magnetic field concentrations at the approximate height of the photosphere, and is represented by panels A1--A4. Here we see two magnetic concentrations of opposite polarities interacting with each other. The time-stamps are separated by 50 s. Approximately 150s into the simulation, one of the polarities cancels out. This is similar to what we observe in Fe~{\sc i} 6302 \AA\ for all events (see panels A1--A4 of Fig~\ref{EB_case3} and Fig~\ref{EB_case8}). The corresponding magnetic field cancellation rate at the photosphere and at the level 430 km above the photosphere is shown in Fig~\ref{muram_fe}. Here, the left panel corresponds to the panels A1--A4 of Fig~\ref{muram_fields}. These magnetic flux curves are plotted by summing the magnetic flux in the opposite polarities, as the polarities evolve in time.   

Furthermore, the intergranular magnetic field concentrations expand into the higher layers of the simulated solar atmosphere due to a magnetic-thermal pressure balance and thermal pressure decrease with height. Such evolution of intergranular magnetic field concentrations are shown in panels B1--B4 of Fig~\ref{muram_fields}. These timestamps are taken at 430 km above the photosphere and are separated by 50 s. The right panel of Fig~\ref{muram_fe} shows a magnetic flux cancellation rate at this level. 

\begin{figure}
\centering
\includegraphics[width=0.45\textwidth, height=0.20 \textheight]{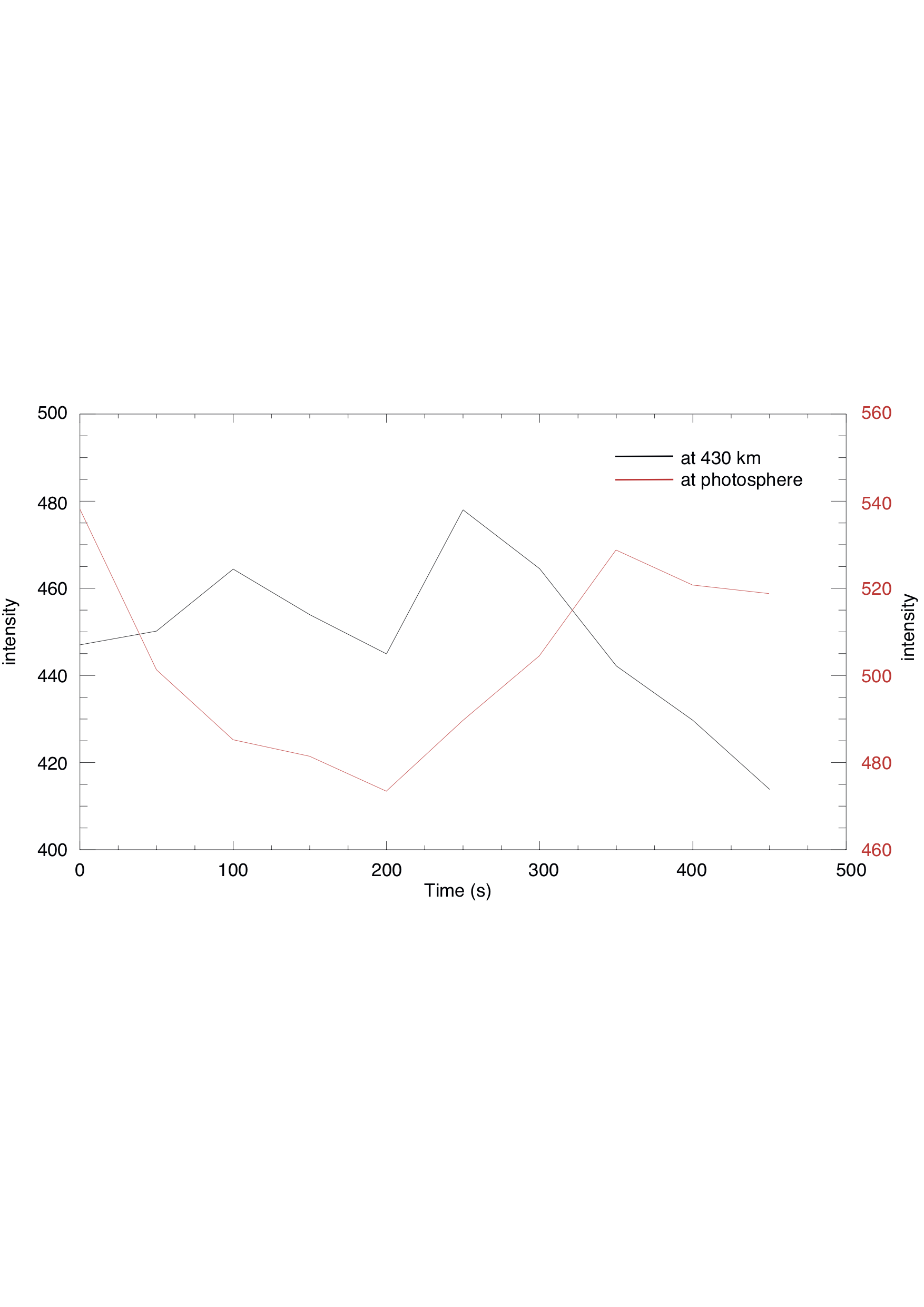}
\caption{Evolution of intensity, corresponding to panels in Fig~\ref{muram_fields} from the MURaM simulations. The timestamps are separated by 50s. The top row with the panels A1--A4, shows temperature change at photospheric levels. The bottom row with panels B1--B2, shows the temperature change at ~430 km above the photosphere.}
\label{muram_temp_lightcurve}
\end{figure}

Due to the geometry of the magnetic field in the simulations, it is expected that the reconnection process evolves in time from the top of the simulation domain towards the solar interior, with the reconnection point moving downwards. This instant of reconnection is seen as an enhancement in the simulations. Fig~\ref{muram_temp}, shows temperature maps taken at the continuum level and lower chromosphere/upper photosphere ($\sim$430 km above the photosphere). Each panel is separated by 50 s. The red arrows show the locations which indicate heating (dark/black colour). Here we see that the temperature rise occurs in panel B1 and continues throughout the evolution. However, in the panels representing the photosphere, the enhancement here is relatively small and appears $\sim$100 s later. Such behaviour matches with the observations. The temperature curves are plotted in Fig~\ref{muram_temp_lightcurve}, the solid red line corresponds to the photospheric continuum level while the solid black line corresponds to the lower chromospheric (upper photospheric) level. The temperature peaks at $\sim$250 s at 430 km at upper photospheric level and at $\sim$350 s for the photosphere.        

\begin{figure}
\centering
\includegraphics[width=0.4\textwidth, height=0.18 \textheight]{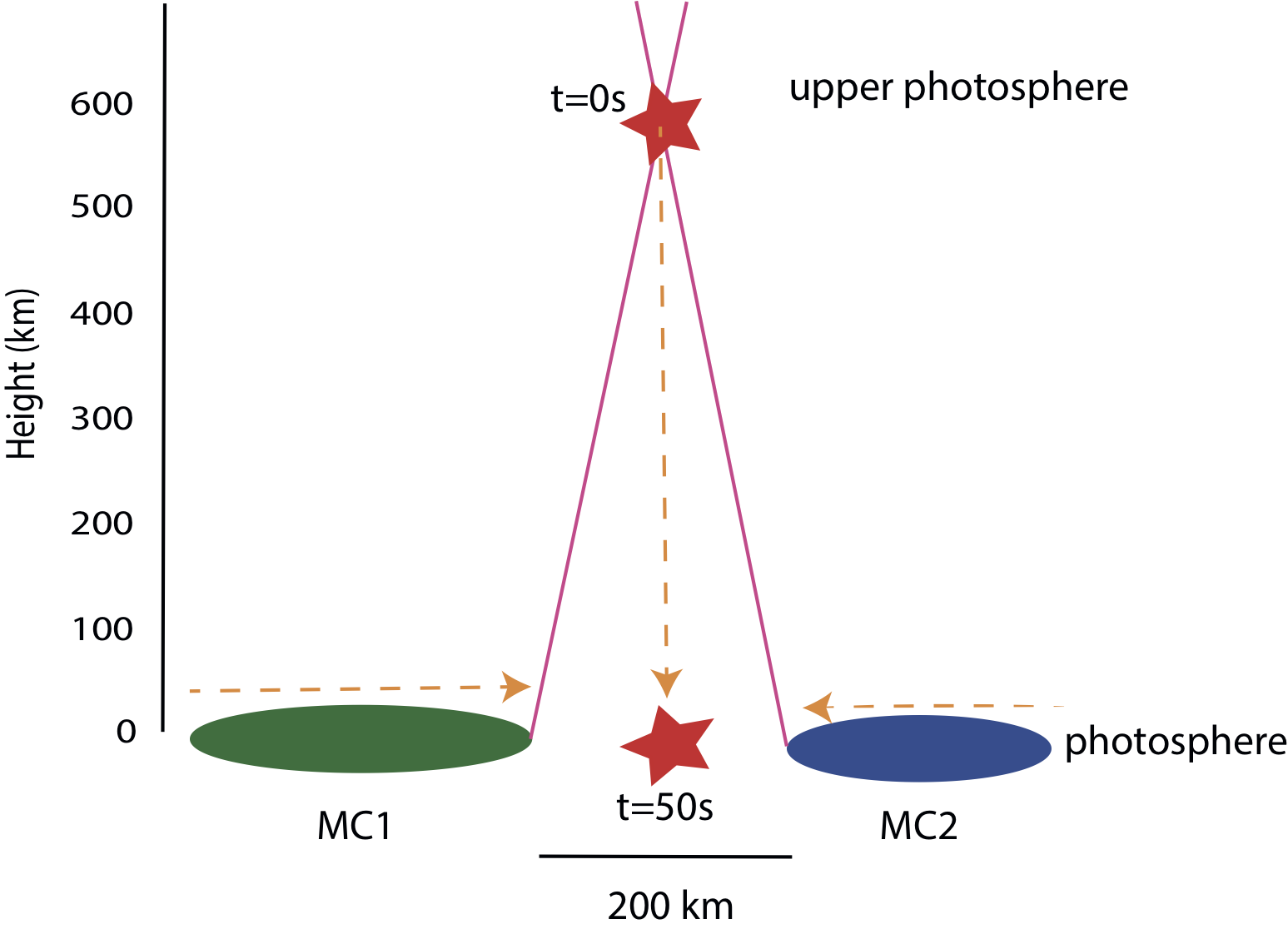}
\caption{Simple cartoon explaining the observed and simulated scenario. Here, we see two MCs separated by 200 km. The arrows indicate the direction of apparent motion. The pink solid lines show the possible straight magnetic field lines. The stars indicate two flaring regions. The flaring at 600 km (upper photosphere) occurs 50s before the flaring at the continuum level. The image is not scaled.}
\label{cartoon}
\end{figure}

The time for the reconnection point to move downwards can be estimated by a similar method described in \cite{2013MNRAS.428.3220K}. Here the authors calculate a velocity of 1.8 km s$^{-1}$ for bright point motions. The mean horizontal speed in the photosphere in the reconnection region, as the simulations show, is $\sim$4 km s$^{-1}$ (2 $\times$ 1.8 km s$^{-1}$). For the sake of simplicity, it can be assumed that the reconnecting magnetic field concentrations move towards each other with this speed. The expansion factor of the intergranular magnetic flux tubes is about 2 between the continuum formation layer and the layer $600~\mathrm{km}$ above it, and the magnetic field concentration size at the photospheric layer is about $200~\mathrm{km}$. Assuming straight field lines, the fields from two flux concentrations touch at $+600~\mathrm{km}$ when there is a $2~\mathrm{km}$ gap between them at the continuum formation layer. It takes about $50~\mathrm{s}$ for the flux concentrations to touch at the continuum formation layer during which time the contact point above has moved down by $600~\mathrm{km}$. Therefore, the time for the reconnection point to move, from the $+600~\mathrm{km}$ layer down to the continuum formation layer is also about $50~\mathrm{s}$. The same calculations work if it takes 100 s to move from the upper photosphere to the continuum layer. The calculation depends on the relative motion speed of the opposite polarity magnetic field concentrations given in Fig \ref{muram_temp_lightcurve}. A simple model for this calculation is sketched in the diagram in Fig \ref{cartoon}. 
 
This is a simplistic calculation, and it does not represent the whole complexity of the dynamic process occurring in the simulation domain. However, it gives us a clue of the scale of the reconnection timescales, and consequently the delay between the upper photospheric and lower chromospheric signals in the Stokes-I profiles of the corresponding absorption lines (e.g. H$\alpha$ 6563 \AA\ and Ca~{\sc ii} 8542 \AA).


\section{Discussions and Summary} \label{diss}

We use observations from H$\alpha$, Ca~{\sc ii} 8542 \AA, and Fe~{\sc i} 6302 \AA\, spectral lines, to investigate the lower solar atmosphere. In Fe~{\sc i} 6302 \AA\ (-40 m\AA), we observe the Stokes-$V$ signal, which is similar to a magnetogram. We observed several events that gave rise to impulsive, flare-like enhancements in the wings of H$\alpha$ and Ca~{\sc ii} 8542 \AA. These events are associated with the interaction of opposite polarities in Stokes-$V$ of Fe~{\sc i} 6302 \AA. The aim of the paper is to show that QSEBs are observed as low intensity contrast events. Fig~\ref{scatterplot}, shows our selected events correspond to a reduction in the Stokes-$V$ intensity accompanied by a maximum H$\alpha$ wing intensity. The peak in the H$\alpha$ wing intensity is $\approx$ 20\% above the average background or less. When compared to the other events in the detection algorithm these events stand out. Thus we have presented QSEBs with less than 20\% intensity increase that satisfies various cancellation models discussed by \citet{2002ApJ...575..506G} and have all the signatures of EBs found in an active region. However, they show a low-intensity impulsive nature (see panels C1--C4 of Fig \ref{EB_case3} and Fig \ref{EB_case8}).  We have used MURaM simulations to understand these events. The sudden enhancement in the wings of the H$\alpha$ line and Ca~{\sc ii} line profiles suggest a physical nature similar to that of EBs. We propose that the reason for the low-intensity contrast of QSEB compared to active region EBs is due to the weaker flux cancellation and the subsequent energy transferred to radiative energy is lower than in regular active region EBs. EBs present near a sunspot have a characteristic recurrent flame-like emission, which recurs with simultaneous H$\alpha$,  and Ca~{\sc ii} 8542 \AA\ wing enhancements. We see such recurring emissions in QSEBs that have a well-defined EB-like morphology (see Fig~\ref{EB_case8}). In addition to the H$\alpha$ signatures, we see an increase in both the core and the wings of the Ca~{\sc ii} line profile. Ca~{\sc ii} profiles associated with the QSEBs are also asymmetrical compared to the H$\alpha$ profile.  
   
Another aspect of our observations is the presence of a temperature increase corresponding to the QSEBs. This temperature increase is especially seen in Ca~{\sc ii} 8542 \AA\ wing emissions. The light-curves (see Fig~\ref{EB_lightcurves_1} and Fig~\ref{EB_lightcurves_2}), show that in most of the cases we see Ca~{\sc ii} 8542 \AA\ wing emissions occurring before the H$\alpha$ wing emissions. The temperature increase further indicates that the increase in emission intensity occurs higher in the upper photosphere and the effects propagate downwards. Such morphology is also observed in the MURaM simulations (see Fig~\ref{muram_temp} and Fig~\ref{muram_temp_lightcurve}). Here we see that the temperature increase occurs higher in the atmospheric layer (at ~7000K or 430km above the photospheric continuum) and occurs before a temperature rise in the photospheric continuum level. This model is further supported by the fact that the SDO channels formed at the continuum level show intensity peaks after the intensity peaks observed in the chromospheric lines of H$\alpha$ and Ca~{\sc ii} 8542 \AA.

Furthermore, our simulations indicates that only a small temperature increase in the lower photosphere is required to reproduce the observed line profiles. This temperature change occurs at the continuum layer 480 km above the assumed photosphere (see Fig~\ref{muram_temp_lightcurve}). These simulation gives us a clue of the scale of the reconnection timescales, and consequently  the delay between the upper photospheric and lower chromospheric signals in the Stokes-I profiles of the corresponding absorption lines (e.g. H$\alpha$ 6563 \AA\ and Ca~{\sc ii} 8542 \AA).






\section*{Acknowledgements}
Armagh Observatory and Planetarium is grant-aided by the N. Ireland Department of Communities. The Swedish 1-m Solar Telescope is operated on the island of La Palma by the Institute for Solar Physics of Stockholm University in the Spanish Observatorio del Roque de los Muchachos of the Instituto de Astrofisica de Canarias. JS was funded by the Leverhulme Trust at Armagh Observatory and now is funded by STFC Grant ST/P000320/1.  JS would like to thank Dr. M. Mathioudakis for financial support from QUB, during the initial development of the project. This work used the DiRAC Data Centric system at Durham University, operated by the Institute for Computational Cosmology on behalf of the STFC DiRAC HPC Facility. DiRAC is part of the National E- Infrastructure. We acknowledge the referee for vital inputs in improving the manuscript. 

\bibliographystyle{plain}
\bibliography{references}


\end{document}